\documentclass[aps,pra,superscriptaddress,twocolumn]{revtex4-2}
\usepackage{amsmath,epsfig,mathrsfs,graphicx,amssymb,color}
\usepackage[T1]{fontenc}
\usepackage{t1enc}
\usepackage{hyperref}
\usepackage{array}
\usepackage{booktabs}

\def\be#1\ee{\begin{equation}#1\end{equation}}
\newcommand{\ba}{\begin{eqnarray} }
\newcommand{\ea}{\end{eqnarray} }

\DeclareMathOperator{\tr}{Tr}

\begin{document}

\title{Null dimension witness based on single measurements}
\author{Josep Batle}
\affiliation{CRISP -- Centre de Recerca Independent de sa Pobla, sa Pobla, Balearic Islands, Spain}
\affiliation{Departament de F\'{\i}sica, Universitat de les Illes Balears, 
                E-07122 Palma de Mallorca, Balearic Islands, Spain}
\email{jbv276@uib.es, batlequantum@gmail.com}
\author{Adam Bednorz}
\affiliation{Faculty of Physics, University of Warsaw, ul. Pasteura 5, PL-02-093 Warsaw, Poland}
\email{Adam.Bednorz@fuw.edu.pl}

%\date{\today}

\begin{abstract}
We present a null witness of the dimension of a quantum system, discriminating real, complex and classical spaces, based on equality due to linear independence.
The witness involves only a single measurement with sufficiently many outcomes and prepared input states.  In addition, for intermediate dimensions, 
the witness bounds saturate for a family of equiangular tight frames including
symmetric informationally complete positive operator valued measures.
Such a witness requires a minimum of resources, being robust against many practical imperfections.
We also discuss errors due to finite statistics.
\end{abstract}

\maketitle

\section{Introduction}

Efficient classical and quantum information technologies rely on few state systems such as bits, qubits and
their networks. Usually, algorithms for error correction and mitigation assume a known dimension of the working systems.
Otherwise, uncontrolled or systematic errors can reduce significantly the accuracy of the realized tasks, accumulated during long sequences. Therefore,
the information processing systems require precise dimension certification.

The dimension certificate should be a numerical criterion based
on the results of a special set of experiments,  a dimension witness. 
The standard construction of the witness is the two-stage protocol, the initial preparation and final measurement \cite{gallego}, 
which are taken from several respective possibilities. The preparation must be independent of the measurement and completed prior to the beginning of the latter.
Initially such witnesses involved
linear and quadratic inequalities, tested experimentally \cite{hendr,ahr,ahr2,dim1}.
However they are not particularly useful when additional contributions from the same space are relatively small. Incidentally, that is the most practical instance, 
taking into account that the physical system is already fabricated as a good approximation of the desired dimension. 
Thus, it is better to use a null witness, i.e. based on equality.
The witness is a functional expression of the results of the experiments, which is exactly zero up to a certain dimension and can be nonzero otherwise \cite{dim,chen}. 

Such a desired witness test is given by the linear independence of the  specific outcome probability $p_{ij}\equiv p(i|j)$ for the preparation $j$ and measurement $i$ by a suitable determinant \cite{dim, chen}. 
In previous works, a witness of dimension $d$ needed $2k$ preparations and $k$ measurements with binary outcomes, with $d\leq k$ 
for the classical system and $d^2\leq k$ for the quantum system \cite{dim}. 
We have recently reduced the number of preparations to $k+1$, preserving the properties of the witness \cite{bb22}.
Experimental null witness tests \cite{dim,bb22} have been demonstrated recently using optical angular momentum and transmon qubits \cite{opt,ibm}.
The disadvantage of the test is to still perform $k$ separate measurements with the total cost of $k\times(k+1)$ experiments.

Here, we show that only a single measurement is necessary, with the number of preparations still being $k+1$. However, the measurements cannot be binary but
must return $k+1$ outcomes. Our witness is zero for $k\geq d,
(d+1)d/2, d^2$ for classical, real, and complex quantum systems, respectively.  
The real quantum system is described by a Hilbert subspace of only real vectors, 
which occur, e.g., when the Hamiltonian is purely imaginary and the unitary operations become real rotations in real space \cite{real}. 
The witness is a determinant of the matrix with entries $p_{ij}$ for the outcome $i$ and the prepared state $j$ and it tests the linear independence of
the underlying Hilbert space.
The measurement operation must be identical and independent of the prepared state in each run of the experiment.

In addition, for the cases when the witness is not zero, one can find an elegant universal bound, 
depending only on the dimension and the number of outcomes and preparations. In fact, the bound can be saturated only for a special family of states
and measurements, equiangular tight frames (ETF) \cite{etf,etf2,etf3,etf4} closely related to a symmetric informationally complete positive operator valued measure (SIC-POVM)
\cite{sic,sic2}. No experiment can measure exactly zero in a finite number of repetitions. Therefore we append the discussion of errors due to finite statistics, unavoidable in each probabilistic test, which are estimated by the adjoint (minor) matrix, for the boundary $k$. 

\begin{figure}
\includegraphics[scale=.3]{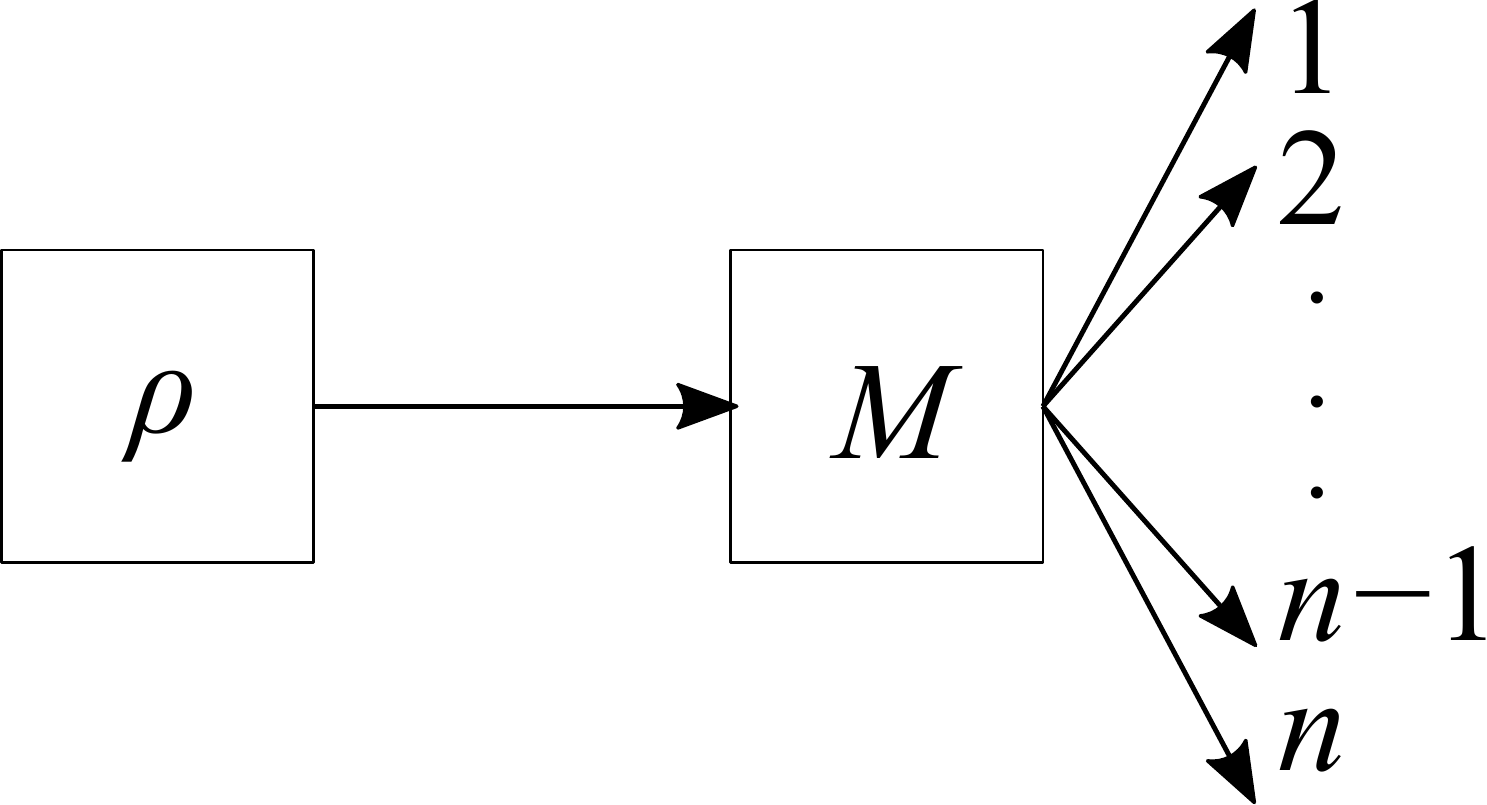}
\caption{Preparation and measurement scenario; the state is prepared as $\rho$ and measured by $M$ to give an outcome  $1\dots n$.}
\label{pms}
\end{figure}

\section{Dimension certificate}
We consider the standard prepare and measure scenario with $n$ outcomes (Fig. \ref{pms}). 
The input state is prepared in one of $m$ possibilities represented by Hermitian matrices 
$0\leq \hat{\rho}_1,\dots,\hat{\rho}_{m}$, $\mathrm{Tr}\hat{\rho}_j=1$.
The probability of the measurement of an outcome $i=1,\dots,n$,  is applied out of $m$ input states,
$p_{ij}=\mathrm{Tr}\,\hat{M}_i\hat{\rho}_j$ for the measurement operators $0\leq \hat{M}_1,\dots,\hat{M}_n$, $\sum_i \hat{M}_i=\hat{1}$.
If the system is classical, with $d$ states,
then $p_{ij}=\sum_{a}q_{ia}r_{aj}$  with $a=1,\dots,d$ and $r$, $q$ describing the transfer probabilities from the prepared state to the classical $d$-dimensional register, whereas $q$ is the transfer probability from the register to the measurement outcome.
The quantum states and measurements can be either real or fully complex.
The real register of a $d-$level quantum state consists of $(d+1)d/2$ Gell-Mann basis matrices with all zeros except a single $1$ on 
the diagonal or a symmetric off-diagonal pair of two $1$'s. The complex register is enlarged by $(d-1)d/2$ antisymmetric matrices with entries $(i,-i)$.

For any dimensions witness, both the number of preparations and outcomes must exceed the dimension $d$.
Otherwise, for $d$ outcomes and $n>d$ preparations, we can generate the desired probability matrix $p_{ij}$ taking $\hat{M}_i=|i\rangle\langle i|$
and $\hat{\rho}_j=\sum_i p_{ij}|i\rangle\langle i|$. On the other hand, having $n>d$ outcomes and $d$ preparations we take 
$\hat{M}_i=\sum_j p_{ij}|j\rangle\langle j|$ while $\hat{\rho}_j=|j\rangle\langle j|$, since $\sum_i p_{ij}=1$. 
Following previous results \cite{dim}, for a $d-$dimensional quantum system we have
\be
\prod_i p_{ii}\leq \mathrm{min}[1,(d/n)^n],\label{inn}
\ee
because the maximum of $p_{ii}$  is bounded by $\mathrm{Tr}\hat{M}_i$ and the geometric mean is bounded by the
average $\sum_i\mathrm{Tr}\hat{M}_i/n=d/n$ \cite{dim}.
A saturating real example is $\hat{\rho}_a=|a\rangle\langle a|$, $\hat{M}_a=\hat{\rho}_ad/n$ and, e.g., Refs.
 \cite{benfic,zimm},
\begin{align}
&\sqrt{d/2}|a\rangle= (2\nmid d)|d\rangle/\sqrt{2}+\nonumber\\
&\sum_{j\leq d/2} [|2j-1\rangle\cos(2\pi aj/n)+|2j\rangle\sin(2\pi aj/n)],
\end{align}
where $j=1,2,\dots$ and $2\nmid d$ means to include this term only if $d$ is odd.

The  classical bound is obtained observing that $M$ becomes a classical Markov process matrix for $d$ states.
The optimal $p_{ii}$ occurs when the state corresponds to the largest element in the $i$th row of $M$. From the pigeonhole principle,
having more outcomes than states, some states must be assigned to multiple outcomes. Again using arithmetic and geometric mean inequality and the fact that the
sum of each column of $M$ is $1$, the optimal case is for this maximum equals $1/q$ for the multiplicity $q$. Therefore the classical bound for (\ref{inn})
is
\be
1/q^{q(d-r)}(q+1)^{(q+1)r}
\ee
if $n=qd+r$ for $r=0,\dots,d-1$. Note that both bounds coincide only when $n$ is the multiplicity of $d$.

The inequality (\ref{inn}) is not satisfactory because (i) it does not discriminate between real and complex quantum systems, (ii) it provides  the same
 classical and quantum bound in some cases, and (iii) it is incapable of detecting small contributions from the extra space.
Therefore we propose a null test of the dimension by taking a single measurement with $n=k+1$ outcomes and preparations.

Let the matrix $p$ be formed with entries $p_{ij}$, the probabilities of the outcome $i$ for the preparation  $j$.
We define the witness 
$
W_k=\det p
$
which is equal to $0$ for $k\geq d,d(d+1)/2,d^2$ for classical, quantum real and quantum complex states of dimension $d$,
because the determinant size exceeds the dimension of spanning the linear space of states. Optionally, the witness can be reduced to
a $k\times k$ matrix noting that the sum of all measurements is equal to $1$, and replacing
\be
p_{ij}\to p_{ij}-p_{i,k+1}.
\ee
The above construction is the main result of the present paper. However, there is another very important inequality, which helps to estimate the scale
of the violation in the above null test by the extra space. \newline

\noindent {\bf Theorem}. The dimension witness $W_k$ satisfies the relation  
\be
W_k\leq \mathrm{min}[1,(d-1)^k/k^k],\label{ine}
\ee
for a system of dimension $d$.\newline

\noindent {\it Proof}. The absolute bound is $W\leq 1$, since $p$ is a Markov process matrix, and therefore from the Perron-Frobenius theorem  all moduli of its eigenvalues are bounded by $1$.
It is saturated for $d>k$ taking $\hat{M}_i=|i\rangle\langle i|=\hat{\rho}_i$. 
Let us then assume $d\leq k$.
We define  $m=\mathrm{Tr}\,\hat{M}$ and $\hat{M}'=\hat{M}/m$ so that $\mathrm{Tr}\,\hat{M}'=1$. Then
\be
\det p=\det p'\prod_i m_i,
\ee
with $\hat{M}_i$ replaced by $\hat{M}'_i$ in $p'$. Since $\sum_i m_i=\sum_i\mathrm{Tr}\,\hat{M}_i=\mathrm{Tr}\,\hat{1}=d$, from the arithmetic and geometric means 
$\prod_i m_i$ has the maximum $[d/(k+1)]^{k+1}$ for $m_i=d/(k+1)$. On the other hand, $\det p'$ is maximized for projective 
$\hat{M}'_i=|m_i\rangle\langle m_i|$ because of the linearity of the determinant with respect to each row separately.
 
\noindent For any normalized set of vectors $u_i$ and $v_i$ such that $p'_{ij}=\langle\langle u_i|v_j\rangle\rangle$,
with some scalar product $\langle\langle\cdot|\cdot\rangle\rangle$, we shall show that the determinant of $p'$
is maximized by either $v_i$ replaced by $u_i$ or $u_i$ replaced by $v_i$.
Let us take the orthonormalization matrix $U$ such that $u_i=\sum_j U_{ij}\tilde u_j$
so that $\langle\langle\tilde u_i|\tilde u_j\rangle\rangle=\delta_{ij}$, e.g., by Gram-Schmidt method (analogously $V$ for $v$ and $\tilde v$).
Then $p'=U^{T}\tilde p'V$
for $\tilde{p}'_{ij}=\langle\langle\tilde u_i|\tilde v_j\rangle\rangle$, and thus
\be
 \det p'=\det\tilde p'\det U\det V.
\ee
By rotation $\tilde u_i=\sum_k O_{ik}e_k$ in some basis $e$ and orthogonal transform $O$, we get
$O^T\tilde p'=p''$ where $p''_{ij}=\langle\langle e_i|\tilde v_j\rangle\rangle$
Now the matrix consists of columns $\tilde v_j$. By virtue of the Hadamard inequality
$
 \det p''\leq \prod_i |v^e_i|,
$
where $|v^e_i|$ is the length vector $\tilde v_i$ projected onto the space $e$ (the original space may be larger). We have $|v^e_i|\leq |\tilde v_i|=1$ and so $\det p''\leq 1$. As $\det O=\pm 1$
we end up with $|\det p'|\leq |\det U\det V|$.
We choose the greater of either $\det U$ or $\det V$ and replace the corresponding set of $v$ by $u$, or vice versa.
 
We apply the  above fact to the Hilbert-Schmidt metric for Hermitian matrices $\hat{u}_i$ and $\hat{v}_i$ with
$\langle\langle u|v\rangle\rangle=\mathrm{Tr}\,\hat{u}^\dag \hat{v}$. Now, the symmetric matrix is positive definite and has real eigenvalues. However, for the vector $a=(1,\dots,1)$
we get $a^Tp'a=\mathrm{Tr}\, \hat{S}^2$ for $\hat{S}=\sum_i \hat{\rho}_i$, with $\hat{\rho}_i=|\psi_i\rangle\langle\psi_i|$.
Then from the Cauchy-Schwarz (CS) inequality $\mathrm{Tr}\, \hat{A}^2\,\mathrm{Tr}\, \hat{B}^2\geq (\mathrm{Tr}\,\hat{A}\hat{B})^2$
applied to $\hat{A}=\hat{S}$, $\hat{B}=\hat{1}$ (the identity), we obtain the Benedetto-Fickus inequality \cite{benfic}
\be
 \mathrm{Tr}\,\hat{S}^2 d\geq (\mathrm{Tr}\,\hat{S})^2=(k+1)^2,
\ee
because $\mathrm{Tr}\,\hat{1}^2=d$, $\mathrm{Tr}\,\hat{S}=\sum_i\mathrm{Tr}\,\hat{\rho}_i=\sum_i 1=k+1$. Therefore $a^TWa\geq  (k+1)^2/d$. Then one of eigenvalues of $W$ must be at least $(k+1)/d$, while the sum of all eigenvalues
 $
 \sum_i \lambda_i=\mathrm{Tr}\,p'=k+1
 $.
Let us recall here that $\det p'=\prod_i \lambda_i$. Let us isolate $\lambda_{k+1}\geq (k+1)/d$. 
From the inequality between arithmetic and geometric means, we retrieve the maximal product when all $\lambda_i=\lambda$. 
Then $\lambda_{k+1}=k+1-k\lambda$ and the determinant reads
$
\lambda^k(1+k-k\lambda)
$
which has the maximum at $\lambda=1$. Therefore the maximum is reached when the other constraint is saturated, namely
$\lambda_{k+1}=(k+1)/d$, giving $\lambda=(1+1/k)(1-1/d)$.
Altogether, $\det p'\leq (k+1)^{k+1}(d-1)^k/d^{k+1}k^k$ leading finally to (\ref{ine}) for $d\leq k$. $\square$\newline

The bound (\ref{ine}) is saturated only when all the inequalities in the proof become equalities. This means that the largest eigenvalue becomes
minimal, i.e. $(k+1)/d$, which occurs only if $\hat{S}=(k+1)\hat{1}/d$ by the property of CS inequality. Moreover, the other eigenvalues must
become equal to saturate the bound due to arithmetic and geometric mean inequality. This means that $|\langle \psi_i|\psi_j\rangle|^2=(k+1-d)/kd$, known as the Welch bound \cite{welch}, 
and $\hat{M}_j=|\psi_j\rangle\langle \psi_j|(k+1)/d$. All these conditions define ETF \cite{etf}. 
Saturation of (\ref{ine}) leads then to the same solution (if it exists) as minimizing the frame potential $\sum_{ij} |\langle\psi_i|\psi_j\rangle|^{2p}$ for $p>1$
\cite{oktay,frpot,frpot2}.
A trivial case occurs when $k=d$ for  states on a regular simplex, $\hat{\rho}_i=|\psi_i\rangle\langle\psi_i|$, 
with $|\psi_i\rangle$ inscribed in a unit sphere, e.g.,
\be
|\psi_i\rangle=\sqrt{1+d^{-1}}|i\rangle-d^{-3/2}(\sqrt{d+1}+1)\sum_j|j\rangle,
\ee
for $i=1,\dots,d$,
$
|\psi_{d+1}\rangle=\sum_j|j\rangle d^{-1/2}
$
and $\hat{M}_i=\hat{\rho}_i d/(d+1)$. 
In particular, for $d=k=2$, the regular simplex is the equilateral triangle
but one can take independent measurement and preparation real bases.
The case $d=2$, $k=3$ is saturated for a regular tetrahedron on the Bloch sphere, i.e., $\sqrt{3}|\psi_j\rangle=|1\rangle+\sqrt{2}\omega^j|2\rangle$, $j=1,2,3$,  
and $|\psi_4\rangle=|1\rangle$
for $\omega=e^{2\pi i/3}=(i\sqrt{3}-1)/2$, and one can take independent (complex) measurement and preparation bases.
Another simple case is $k=5$, $d=3$ for the vertices of a regular icosahedron.

Unfortunately, there is no straightforward method to determine if there exists an ETF for a given pair $(d,k)$. We refer the reader
to the mathematical literature discussing and tabularizing a large number of cases, involving nontrivial algebraic properties 
\cite{etf,etf2,etf3,etf4} (see also the Appendix).
From a physical point of view, it is relevant that the maximal $k+1$ for a given $d$ for the possible ETF is $d(d+1)/2$ and $d^2$ in the real
and complex case, respectively. However, the real case does not always provide ETF (e.g., for $d=4$) but the complex maximal ETF (SIC-POVM)
 is conjectured to exist in all dimensions \cite{zaun,sic,sic2}. 

\begin{table}
\begin{tabular}{*{12}{c}}
\toprule
$k$&2r&2c&3r&3c&4r&4c&5r&5c&6r&6c\\
\midrule
1&{\bf 1}&&&&&&&&&\\
2&{\bf 0.25}&&{\bf 1}&&&&&&&\\
3&0&{\bf 0.037}&{\bf 0.296}&&{\bf 1}&&&&&\\
4&0&0&0.053&0.059&{\bf 0.316}&&{\bf 1}&&&\\
5&0&0&{\bf 0.010}&&0.073&0.075&{\bf 0.328}&&{\bf 1}&\\
\bottomrule
\end{tabular}
\caption{The quantum maximum of $W_k$ for the $(k+1)\times (k+1)$ matrix for a $d$-dimensional system, either real (r) or complex (c).
An empty cell means the value is the nearest number to the left in the row. The bold value saturates (\ref{ine})}
\label{taboneq}
\end{table}

Apart from the cases saturating (\ref{ine}), 
the search for the maximum of the witness for each pair $(d,k)$ is hard if tackled only from the analytical point of view, 
although it is always in principle some algebraic number. The numerical results are presented in Table \ref{taboneq}.  
Our search is somewhat similar to the search of ETF and SIC-POVM, which also could not avoid numerical calculations
except for some special cases \cite{sic2}. 
The optimization is taken over the states $\hat{\rho}_1,\dots,\hat{\rho}_{k+1}$, represented by pure states $\hat{\rho}_j=|\psi_j\rangle\langle\psi_j|$
(they are optimal by linearity of the determinant with respect to each columns).
The vectors $|\psi_j\rangle$ are real or complex unit vectors in $\mathbb{R}^d$ and $\mathbb{C}^d$.

The outcome operators $\hat{M}_i$ are not fully independent as they are positive and must be summed to $\hat{1}$.
Construction of the optimal $\hat{M}_i$ goes as follows. 
Without loss of generality, suppose that $\hat{M}_1$ and $\hat{M}_2$ have rank $>d/2$. Then their support spaces overlap, i.e. there exists a nonzero 
$\hat{M}_{0}\leq\hat{M}_{1,2}$. From linearity, we can increase the witness, moving the whole $\hat{M}_0$ to either of $\hat{M}_{1,2}$.
Therefore we can assume rank $\leq d/2$.
For the numerical analysis,
we simply generate any positive Hermitian matrix, e.g., by a combination of projections in some basis, and rotate and scale them.
In particular,
$\hat{M}'_i=\hat{S}^2_i$ where $\hat{S}_i$ is a diagonal matrix with entries $\sin\phi_{ij}$, $j\leq \lfloor d/2\rfloor$ and $0$ otherwise, 
while $\hat{C}_i$ is diagonal with entries $\cos\phi_{ij}$ for $j\leq \lfloor d/2\rfloor$ and $1$ otherwise.
Then we correct
\be
\hat{M}_{i+1}=\hat{C}_{1}\hat{U}_{1}\cdots \hat{C}_i\hat{U}_i \hat{M}'_{i+1}\hat{U}^\dag_i\hat{C}_i\cdots \hat{U}^\dag_1\hat{C}_1,
\ee
for all $i<k$ while $\hat{M}_1=\hat{M}'_1$ and $\hat{M}_{k+1}=\hat{1}-\sum_{i\leq k} \hat{M}_k$. 
Here $\hat{U}_i$ are arbitrary orthogonal/unitary rotations in the real/complex case.
With a proper parametrization, the problem consists of finding the supremum of $W_k$.  
The details of such algorithms are given in Ref. \cite{bb22} while the results are presented in the Appendix. 
In all explored cases, we found the optimal cases for $\hat{M}_i$ of rank 1 and conjecture that this is a general property.

In the special case  $k+1=(d+1)d/2$ and $k+1=d^2$ for  real and complex quantum states of dimension $d$, respectively,
it is equal to the dimension of the operator space spun by the appropriate Gell-Mann matrices.
Then the determinant can be written as a product of determinants of separate square matrices for the preparations $\hat{\rho}_j$ and outcome measurements
$\hat{M}_i$ so the maximum can be determined separately for $\hat{M}$ and $\hat{\rho}$ and the same maximum is reached if preparations  
are rotated by an arbitrary orthogonal/unitary matrix $\hat{R}$/$\hat{U}$ (the same for all preparations) in the real/complex case, i.e., $\hat{\rho}_j\to \hat{R}\hat{\rho}_j\hat{R}^T$ or
$\hat{\rho}_i\to\hat{U}\hat{\rho}_i\hat{U}^\dag$.

\section{Extra space detection}

The witness is useful to detect higher dimensions of the system by checking deviations from $0$ for $k= d^2$ or $d(d+1)/2$.
Let us suppose that the system is designed to be a perfect $d$-level state but there is a small contribution from the extra space $\delta \hat{\rho}$
and $\delta\hat{M}$. This deviation is given by $\delta p_{ij}=\tr \delta \hat{M}_i \delta\hat{\rho}_j$.
Then the small contribution from the extra space reads, due to the Jacobi identity,
\be
\bar{W}\simeq W+\tr \delta p\; A\label{dev}
\ee
where $A=\mathrm{Adj}\; p$ is the adjoint matrix to $p$. Note that we cannot use the fact $A=p^{-1}\det p$
as $\det p=0$.

However, even without extra space the null prediction concerns an asymptotic case in the limit of infinite series of identical and independent experiments.
No experiment will produce an exact zero in finite statistics. In fact, the null hypothesis remains valid if 
the witness is within the error bound, determined  mainly by the number of trials. Suppose $N$ trials for each preparation.
Then $\bar{p}_{ij}=n_{ij}/N$ where $n_{ij}$ is the number of outcomes $i$ occurring in $N$ trials with the preparation $j$.
It is then a random variable and we can denote its deviation from the asymptotic limit $p_{ij}$ as $\delta p_{ij}=\bar{p}_{ij}-p_{ij}$.
The witness also becomes a random value $\bar{W}$ depending on the actual results of the experiments, $\bar{p}$.
We shall estimate average $\langle\bar{W}\rangle$ and deviation $\langle(\delta \bar{W})^2\rangle$ for $\delta\bar{W}=\bar{W}-\langle\bar{W}\rangle$.
Expanding $\bar{p}=p+\delta p$ for small $\delta p$ we can use (\ref{dev}) to estimate possible errors.
Then $\langle \delta p\rangle=0$ and $N\langle \delta p_{ij}\delta p_{qr}\rangle=\delta_{jr}(p_{ij}\delta_{iq}-p_{ij}p_{qj})$
which gives $\langle\bar{W}\rangle=W=0$ and
\be
N\langle(\delta\bar W)^2\rangle=\sum_{ij}p_{ij}(A_{ji}-\bar{A}_j)^2=\sum_{ij}p_{ij}(A_{ji}^2-\bar{A}_j^2)
\ee
for $\bar{A}_j=\sum_i p_{ij} A_{ji}$, taking into account $\sum_i p_{ij}=1$. A similar error analysis has been successfully applied in the real experiments
\cite{opt,ibm}.

There is no \emph{a priori} optimal set of measurements and the witness is expected to remain zero always.
However, one should avoid a deliberate reduction of the space, e.g., taking real instead of complex space.
In principle one can also use higher $k$ but it is less reliable as (a) the adjoint matrix becomes zero and so the errors
need second-order minors, and (b) at such small errors one should also discuss the instability of the probability distribution due to the calibration drifts.
This is beyond the scope of this paper as they violate the assumptions of our test. Nevertheless, it would need a modification of the test, taking into account
a reasonable model of instabilities.

\section{Conclusion}
Our null dimension witness based on a single measurement can speed up diagnostic tests of qubits and other basic
working systems of quantum computers. A many-outcome measurement can be implemented coupling a single qubit to, e.g., auxiliary ones. The witness can be generalized, modifying the assumptions, to include compound systems (e.g., two qubits
with controlled interactions), or weakening the independence condition (e.g., imposing only partial restrictions). It is also important to
determine the possible physical sources of potential extra space, e.g., higher excitations.
In any case, dimension diagnostics 
will remain essential for successful quantum error correction and mitigation.

\section*{Acknowledgements}
 We are grateful to J. Tworzyd{\l}o and S. So{\l}tan for inspiring discussions and W. Bednorz for checking the mathematical proof presented in this work. J.B. acknowledges fruitful discussions with 
Maria del Mar and Regina Batle.

\appendix
\section{Numerical search for the maximum and equiangular tight frames}

We shall summarize the algebraic and numerical search for the maximum of $W_k$.
For $d=3$, $k=4$ the maximum is $(3/8)^3=0.052734375$ in the real case for $2|\psi_{1,2}\rangle=|1\rangle\pm\sqrt{3}|2\rangle$ and
$2|\psi_{3,4}\rangle=|1\rangle\pm\sqrt{3}|3\rangle$, $|\psi_5\rangle=|1\rangle$, with $2\hat{M}_j=|m_j\rangle\langle m_j|$ for $|m_{1,2}\rangle=|1\rangle/2\pm |2\rangle$, $|m_{3,4}\rangle=|1\rangle/2\pm |3\rangle$, $|m_5\rangle=|1\rangle$.
 
 For the complex case, the maximum is numerical for a special class of states and measurements, namely
$|\psi_j\rangle=|j\rangle$, and
 $M_j=C^2|\psi_j\rangle\langle \psi_j|$ for $j=1,2$ while
 $\hat{M}_j=|m_j\rangle\langle m_j|$ and
 \begin{eqnarray}
 \sqrt{2}|\psi_j\rangle&=&A(\omega^j |1\rangle+\omega^{2j}|2\rangle)+\sqrt{2}B|3\rangle, \cr
 \sqrt{3}|m_j\rangle&=&D(\omega^j|1\rangle+\omega^{2j}|2\rangle)+|3\rangle,
 \end{eqnarray}
 for $j=3,4,5$ and $\omega=e^{2\pi i/3}$. Then, for the constraints $A^2+B^2=C^2+D^2=1$, the determinant reads
 $$
 C^4A^2B^2D^2(\sqrt{2}B+AD/2)^2,
 $$
 which is approximately $0.0585806$ at $A,D=0.617344,0.603104$. It is still below (5), giving $1/16=0.0625$.
 
 Another saturating case is
 $d=3$, $k=5$ for the vertices of a regular icosahedron \cite{equi},
 \be
\sqrt{\phi+2}|\psi_{s+j}\rangle=\phi|j\rangle+(-1)^s|j+1\rangle,
\ee
for $|j+3\rangle\equiv |j\rangle$, the golden ratio $\phi=(1+\sqrt{5})/2$, $s=0,3$,
and measurements $\hat{M}_j=|\psi_j\rangle\langle\psi_j|/2$. The real bases of the measurement and preparations are independent.

In the real case $k=5$, $d=4$, the optimal value is attained for the
states 
\be
|\psi_{s+j}\rangle=a|j\rangle+(-1)^s b|4\rangle,
\ee
for $j=1,2,3$, $s=0,3$,
with  $a=\sqrt{7/10}$, $b=\sqrt{3/10}$,
and measurements $\hat{M}_j=|m_j\rangle\langle m_j|$
\be
|m_{s+j}\rangle=A|j\rangle+(-1)^s B|4\rangle,
\ee
for $A=1/\sqrt{2}$ and $B=1/\sqrt{6}$ giving $W=7^3\sqrt{7}/4\cdot 5^5 \simeq 0.07259941597561233$.
 
In the complex case, the generic optimal family reads
\begin{align}
&|\psi_j\rangle=x|1\rangle+y|2\rangle+z\omega^j|3\rangle+t\omega^{2j}|4\rangle,\nonumber\\
&|\psi_{j+3}\rangle=y|1\rangle+x|2\rangle+t\omega^j|3\rangle+z\omega^{2j}|4\rangle,
\end{align}
and $\hat{M}_j=|m_j\rangle\langle m_j|/3$
with
\begin{align}
&|m_j\rangle=|1\rangle+A\omega^j|3\rangle+B\omega^{2j}|4\rangle,\nonumber\\
&m_{j+3}\rangle=|2\rangle+B\omega^j|3\rangle+A\omega^{2j}|4\rangle,
\end{align}
and real $A,B,x,y,z,t$,  $A^2+B^2=x^2+y^2+z^2+t^2=1$ (circle and  3-sphere).
Then the generic determinant
\begin{align}
&(x z A - y t A - y z B + x t B)^2\times\nonumber\\
& (x z A + y t A + y z B + x t B + 2 z t A B)^2\times\nonumber\\
& (x^2 - y^2 + (z^2 - t^2)(A^2 - B^2))
\end{align}
gives the numerical maximum $\simeq 0.074847$, below our universal bound $0.07776$.

For a comprehensive list of ETFs, see Ref. \cite{etf}. Here we summarize the most regular cases.
In the complex case one can take a SIC-POVM for $k=d^2-1$ with independent measurement and preparation bases \cite{sic2}.
Moreover, for a prime $p$ such that $4|p+1$ (e.g., $p=3,7,11,19,23$), $k=p-1$, and 
\be
|\psi_j\rangle=\sum_{a=0}^{d-1}\zeta^{ja^2}|a+1\rangle,
\ee
for $d=(p+1)/2$, or
\be
|\psi_j\rangle=\sum_{a=1}^{d}\zeta^{ja^2}|a\rangle,
\ee
for $d=(p-1)/2$,
with $\zeta=e^{2\pi i/p}$, because of the Gaussian quadratic identity $\sum^{(p-1)/2}_{a= 0} \zeta^{ja^2}=\pm i\sqrt{p}$ for $j<p$ \cite{kal}.
For all odd primes $p$, $\sum^{(p-1)/2}_{a= 0} \zeta^{ja^2}=s\sqrt{p}$  with $s^4=1$, which gives another example for $d=(p+1)/2$, $k=p$ with
\cite{zaun}
\be
\sqrt{p}|\psi_j\rangle=|d\rangle+\sum_{a=0=1}^{d-1}\sqrt{2}\zeta^{ja^2}|a+1\rangle,
\ee
for $j=1,\dots, p$ and $|\psi_{p+1}\rangle=|d\rangle$.
Other real examples are, e.g., known equiangular sets of lines \cite{equi} (see also direct constructions \cite{trem,trem2}).
By completion to the orthonormal basis one can show duality, i.e., if there exists an ETF $(d,k)$ then it exists also for $(k+1-d,k)$ \cite{zaun,etf4}.
Expanding $|\psi_j\rangle=\sum_a \psi_{aj}|a\rangle$ from 
\be
(k+1)\hat{1}/d=\sum_j|\psi_j\rangle\langle\psi_j|=\sum_{abj}\psi_{aj}|a\rangle\langle b|\psi_{bj}^\ast
\ee
we get $\sum_{abj}\psi_{aj}\psi^\ast_{bj}=(k+1)\delta_{ab}/d$. Treating the sequence $\psi_{a1},\dots,\psi_{a\,k+1}$
as entries of the vector $\Psi_a$ we see that they are orthogonal. By scaling with $\sqrt{d/(k+1)}$ they become normalized $\Psi'_a$
and we can complete the set by the remaining orthonormal $\Psi'_a$ for $a=d+1,\dots,k+1$. Rescaling the latter by $\sqrt{(k+1)/(k+1-d)}$
we get the entries $\psi_{aj}$ for the dual ETF as orthogonality works in both directions, with
\be
\sum_a \psi'_{ai}\psi^{\prime\ast}_{aj}=\delta_{ij}.
\ee
Since $\sum_a=\sum_{a\leq d}+\sum_{a>d}$ and the former is $\sqrt{(k+1-d)/dk}d/(k+1)=\sqrt{(k+1)/k(k+1-d)}(k+1-d)/(k+1)$ for $i\neq j$,
it satisfies all the conditions of ETF.

\end{document}